\documentclass[aip,graphicx,reprint]{revtex4-2}

\usepackage{graphicx,bm,amssymb,amsmath}
\usepackage{color}
\usepackage{multirow,tabularx,diagbox,slashbox}

\draft 

\begin{document}

\title{Free energy of self-avoiding polymer chain confined between parallel walls} 



\author{M\'arcio S. Gomes-Filho}
\affiliation{Centro de Ciências Naturais e Humanas, Universidade Federal do ABC, 09210-580, Santo André, São Paulo, Brazil}
\affiliation{ICTP South American Institute for Fundamental Research,\\ Institute of Theoretical Physics, São Paulo State University, Rua Dr. Bento Teobaldo Ferraz, São Paulo 01140-070, SP, Brazil.}

\author{Eugene M. Terentjev}
\email{emt1000@cam.ac.uk}
\affiliation{Cavendish Laboratory, University of Cambridge, JJ Thomson Avenue, Cambridge CB3 0HE, U.K.}

\date{\today}

\begin{abstract}\noindent Understanding and computing the entropic forces exerted by polymer chains under confinement is important for many reasons,  from research to applications.
However, extracting properties related to the free energy, such as the force (or pressure) on confining walls, does not readily emerge from conventional polymer dynamics simulations due to the entropic contributions inherent in these free energies.
Here we propose an alternative method to compute such forces, and the associated free energies, based on empirically measuring the average force required to confine a polymer chain between parallel walls connected by an artificial elastic spring. This measurement enables us to interpolate the expression for the free
energy of a confined self-avoiding chain and offer an analytical expression to complement the classical theory of ideal chains in confined spaces.
Therefore, the significance of our method extends beyond the findings of this paper: it can be effectively employed to investigate the confinement free energy across diverse scenarios where all kinds of polymer chains are confined in a gap between parallel walls.
\end{abstract}


\maketitle 
\section{Introduction}

Understanding forces, or the pressure exerted by polymer chains within confined spaces is a fundamental problem in polymer science. This comprehension is demanded across diverse domains, including biological processes, nanotechnology, drug delivery, and microfluidics, among others, underscoring its relevance in both fundamental research and technological applications~\cite{Cifra23, Richter19}. For example, it sheds light on how proteins fold under confinement conditions~\cite{mittal2008, taylor2017}, the forces required to package (or eject) biopolymers (e.g., DNA) into (or from) a bacteriophage capsid~\cite{kindt2001, ben2013}, the dynamic properties and conformation of confined DNA~\cite{Dong-qing22, bonthuis2008, tang2010, reisner2012}, and polymer translocation through narrow channels~\cite{muthukumar2016, huang2019, Seth2020}.

Many of these intriguing phenomena are associated with spatial constrains, which emerge from physical boundaries like membrane walls or channel boundaries. These constraints reduce the number of allowed configurations, resulting in a decrease of the conformational entropy and generate a corresponding free energy excess. As a result, the confinement free energy of a polymer is primarily  determined by the entropic effects~\cite{cacciuto2006, sakaue2006,smyda2012}.

Commonly, three length scales can be used to characterize the confinement regimes: the size of the unconfined polymer (measured, for instance, by its radius of gyration $Rg$), the persistence length $l_p$, and the confinement length scale $d$. For instance, the strong confinement regime is defined when $d < l_p$, when $l_p < d \ll 2R_g$ the confinement is considered moderate, while the weak confinement regime corresponds to $d > 2R_g$~\cite{gorbunov1995,smyda2012, leith2016}.

In this context, a polymer chain confined in a gap between two parallel walls  becomes a classical and fundamental problem that exposes the essential physics. The exploration of this scenario traces back to seminal works by Casassa~\cite{casassa1967}, and Edwards and Freed~\cite{edwards1969}, which addressed the excess free energy of an ideal (Gaussian) chain under such confined conditions. Since then, significant progress has been made in this field, employing diverse theoretical approaches,  as well as basic scaling arguments~\cite{DeGennes, gorbunov1995, sakaue2006, milchev2011, MICHELETTI2011, Taylor22}. 

For example, the classical analytical theory has been developed for the excess free energy of an ideal chain ($N$ units of size $\sigma$) confined between two parallel walls at a distance of $d$ (confinement length) within a moderate confinement regime~\cite{casassa1967, edwards1969}. This classical Edwards theory predicts a confinement free energy scaling of $F \sim 1/d^2$. Consequently, the corresponding repulsive entropic force $f = - \partial F/ \partial d$ exerted on the walls scales as $f \sim 1/d^3$. In contrast, in a weak confinement regime, the confinement free energy follows the scaling $F \sim 1/d$, leading to repulsive force scaling as $f \sim 1/d^2$.\cite{gorbunov1995}
On the other hand, the theory for a polymer in good solvent (the self-avoiding chain) becomes more complicated due to the pair interactions (excluded-volume)~\cite{flory1953,edwards1965,DoiEdwards}. {Much effort has also been made to investigate dilute solutions of flexible polymer chains in a good solvent confined between two parallel repulsive walls using field-theoretic methods.\cite{Schlesener2001, Romeis2009}}  

Under confinement, the de~Gennes scaling predictions~\cite{DeGennes} are commonly employed in the literature, and the basic Flory theory can be rewritten in terms of geometrically confined space. For instance, in the case of a self-avoiding chain confined between two walls, this becomes~\cite{milchev2011}:
 \begin{equation} \label{eq:flory}
  \frac{\Delta F}{k_BT} = \frac{R_{||}^2}{N\sigma^2} + \frac{\sigma^{3} N^2 }{d R_{||}^2},
 \end{equation}
where the first term  stands for the usual the (Gaussian) entropic elasticity of the chain with the end-to-end distance $R_{||}$, while the second term accounts for the contribution  of excluded volume interactions in the `pancake' volume of thickness $d$. In this context,  $R_{||}$ becomes the lateral chain dimension in the plane without geometric constrains (for more details see \cite{milchev2011, paturej2013, ha2015}). Note that the `standard' minimization with respect to the lateral size $R_{||}$ gives the equilibrium free energy  scaling with the confinement length as $1/d^{1/2}$, and so the repulsive force scaling as   $1/d^{3/2}$.
Within the blob scaling theory, the free energy of a polymer chain confined in narrow space (a flat slit or a narrow tube) is given by: $F \sim 1/d^{1/\nu}$ with the `Flory exponent' $\nu$ reflecting the chain nature. It should be noted that for the ideal chain ($\nu = 1/2$) one recovers the analytical result of Edwards and Freed for the ideal chain confined in narrow space. For the confined self-avoiding chain: $F \sim 1/d^{1.7}$.~\cite{DeGennes}  It would be important to be able to verify (or question) these analytical results with an appropriate computer simulation, which is what we aim for in this paper. 

Technically, today it is easy to carry out {dynamic} simulations of a chain in confined space~\cite{milchev2011, ha2015} and determine its statistical parameters such as $R_g$, but it is not straightforward to find the free energy-related properties, such as the force (or the pressure) on confining walls. This is because the entropic contribution to these free energies, and the associated forces, are not naturally coming out of a typical time-limited computer simulation.

{Although various advanced methods exist for estimating free energy in Molecular Dynamics (MD) simulations,\cite{bonomi2009, frenkel2001, Parrinello2002, kumar1992} 
accurately accounting for entropy-related contributions remains a fundamental challenge. Addressing this difficulty has been a key focus of many studies,  including those by Frenkel (1984),\cite{frenkel1984,frenkel2001} Jarzynski (1997),\cite{jarzynski1997} and Parrinello (2002).\cite{Parrinello2002} In this manner, modern approaches, such as thermodynamic integration and umbrella sampling, provide effective strategies for computing free energy in MD simulations.\cite{frenkel2001} However, their application to confined polymer chains remains relatively unexplored.}

{In contrast, Monte Carlo (MC) simulations have been more widely used to estimate the confinement free energy of polymer chains~\cite{milchev1998polymer, cifra1999, cifra2001, hsu2004:jcp, Wang2004, smyda2012, cacciuto2006, Cifra23}. However, a major limitation of standard fixed-length MC simulations is their inability to directly estimate the free energy and the  associated confinement forces.\cite{hsu2003, dickman1998} To adress this challenge, advanced sampling techniques, such as the Pruned-Enriched Rosenbluth Method (PERM) have been employed to enhance the exploration of the conformational space in confined polymer systems~\cite{grassberger1997pruned}. A recent application includes the estimation of the free energy of semiflexible polymers under confinement, covering different confinement regimes.\cite{Tree2014, Cheong2018:pre}}

As an alternative,  recent  studies using Brownian Dynamics (BD) simulations~\cite{dimitrov2008, leith2016} have computed the average force produced by the confined polymer on the walls, and then the corresponding free energy could be estimated by integration.  In particular, Leith et al.~\cite{leith2016} estimated the free energy of a semiflexible polymer chain confined in a slit (which is exactly our problem here) using   Monte Carlo (MC) and BD simulations for weak and strong confinements, respectively.
For moderate confinement regimes, both approaches were in agreement with each other. Within BD simulations, they measured the average force acting on the walls from measuring the potential energy due to the monomer-wall  interactions (repulsive Lennard-Jones potential). The free-energy was then obtained via numerical integration of the average force over the confinement length. By adding a numerical constant obtained from an empirical relation to the free energy,  they found good  results, which turned out to be in agreement with scaling predictions.

However, it is important to note that although the average force exerted by a confined polymer chain can, in principle, be estimated  in a typical simulation through changes in the potential energy of pair interactions, such a force has its physical meaning different from the actual entropic force associated with the confinement free energy based on the change in the number of conformations. As the chain tries to avoid the confinement, producing a force on the walls in order to maximize the number of allowed configurations, its configurational entropy changes. 

In light of this, here we propose an alternative and very simple simulation method based on empirically measuring the average force required to confine the chain.
This measurement enables us to interpolate the expression for the free energy of a confined self-avoiding chain, by integrating the measured force, and thus establish a closed analytical expression to complement the classical theory of polymers in confined space. Therefore, the significance of our method extends beyond the findings of this paper: its concept can be effectively employed to investigate the confinement free energy across diverse scenarios, for example, for different types of chains, and for different types of confinement, as long as the moveable walls controlled by a spring force are constructed.

\section{ Computational details}\label{sec:model}

In this work, we use the currently `standard' approach of the simulation package Large-scale Atomic/Molecular Massively Parallel Simulator (LAMMPS)\cite{lammps01,lammps02} to perform Brownian Dynamics simulations\cite{grest1986:pra, kremer1990:jcp, lappala2013:macro}. It involves numerically integrating the Langevin equation for all interacting particles within the system, enabling us to observe the stochastic temporal evolution of the system. As a result, the system reaches thermal equilibrium at a specified target temperature by applying the Langevin thermostat, which is implemented in LAMMPS within the framework of classical molecular dynamics. The fixed number of particles $N$ connected along the polymer chain within a defined volume~$V$ is kept constant during the simulation. Since this is widely used by many authors, we shall be brief in describing the relevant simulation details.

We use the classical Kremer-Grest bead-spring model for polymers~\cite{grest1986:pra, kremer1990:jcp}, in which $N$ beads (monomers) are connected along the polymer chain through a non-harmonic spring model, made of an attractive finite extensible nonlinear elastic (FENE) potential:
\begin{equation}
{U^{\mathrm{ch}}(r)=}
\begin{cases}
  -\frac{1}{2}\kappa R_{0}^2 \ln \left[ 1 -  \left(\frac{r}{R_{0}} \right)^2 \right],  \hspace{0.5cm}  r \leq R_0    \nonumber \\
   \infty,  \hspace{0.5cm}  r > R_0,  \label{eq:fene}
\end{cases}
\end{equation}
 and an added repulsive truncated Lennard-Jones potential (also referred as Weeks-Chandler-Andersen (WCA) potential~\cite{weeks1971:jcp}) cut off at $r = 2^{1/6}\sigma$:
\begin{equation}
{U(r)=}
\begin{cases}
       4\epsilon^* \left[ \left(\frac{\sigma}{r} \right)^{12} -  \left(\frac{\sigma}{r} \right)^{6} + \frac{1}{4} \right],   \hspace{0.5cm}  r \leq 2^{1/6}\sigma   \nonumber  \\
       0,   \hspace{0.5cm} r > 2^{1/6}\sigma,      \label{eq:lj-fene}
\end{cases}
\end{equation}
where $r$ is the center-to-center distance between consecutive  beads,  $\epsilon^*$ is the repulsive LJ  strength  and $\sigma$  the diameter of an individual monomer.

The FENE potential exhibits harmonic behavior around its minimum. The spring constant is defined as $\kappa = 30\epsilon^*/\sigma^2$. Additionally, the polymer chain extension is constrained, preventing it from stretching beyond the maximum bond length of $R_0 = 1.5\sigma$. These parameter values were chosen in alignment with other computational studies~\cite{kremer1990:jcp, auhl2003, liu2019, Lappala2019}.

The bending stiffness of the polymer chain is introduced through the bending elasticity energy on each bond, which is  given by:
 \begin{equation}
    U^{\mathrm{stiff}}(\theta)=K_{\theta}( 1 + \cos\theta ),
 \end{equation}
where $\theta$ is the angle formed between two consecutive bonds and  $K_{\theta}$ is the bending coefficient.

The interaction between non-bonded particles in the polymer chain can be described by the standard Lennard-Jones (LJ) potential:
\begin{equation}
{U^{\mathrm{LJ}}(r)=}
\begin{cases}
       4\epsilon \left[ \left(\frac{\sigma}{r} \right)^{12} - \left(\frac{\sigma}{r} \right)^{6}\right] - \phi,  \hspace{0.5cm} r \leq r_{\mathrm{cut}} \nonumber \\
       0,    \hspace{0.5cm}   r > r_{\mathrm{cut}}, \label{eq:lj}
\end{cases}
\end{equation}
where $r$ represents the center-to-center distance between beads, $\epsilon$ denotes the depth of the LJ potential well, and the constant $\phi \equiv U^{\mathrm{LJ}}(r_{\mathrm{cut}})$ ensures that $U^{\mathrm{LJ}} \to 0$ as $r \to r_{\mathrm{cut}}$. In this way, the LJ potential can be employed to account for excluded-volume interactions between monomers and to incorporate long-range attraction interactions if and when required. While the full LJ potential describes conditions of poor solvent, when the effective attraction between monomers occurs, a purely repulsive LJ potential represents a good solvent. This repulsive potential is achieved by truncating the LJ potential at its minimum value, corresponding to $r_0 = 2^{1/6}\sigma$~\cite{ceperley1978, lappala2013:macro, lappala2015:macro}.

Consequently, the polymer model described above allows for the simulation of both a self-avoiding chain (with excluded-volume interactions) and an ideal phantom (Gaussian) chain that can intersect itself due to the absence of self-repulsion (without excluded-volume interactions), with the non-bonded LJ potential turned off. This does not have to be the case, and more complex pair interactions can be explored, but here we aim to preserve the ultimate simplicity of the polymer model to illustrate the method of force calculation most clearly.

Unless otherwise specified, the average temperature was kept constant at $k_{\mathrm{B}} T = 1.0 \epsilon$.  The internal to LAMMPS damping constant $damp$ and the LJ time unit were linked to the same energy scale via the fluctuation-dissipation theorem:  $damp = 0.5\tau^{-1}$ and  $\tau = \sigma\sqrt{m/\epsilon}$, where $m$ denotes the bead mass~\cite{grest1986:pra, kremer1990:jcp}. Another important parameter is the  bending coefficient
 $K_{\theta}$,  which is related with the persistence length, $l_{\mathrm{p}}=\sigma(K_\theta/k_{\mathrm{B}}T)$, and for our basic case of a flexible polymer chain, we set $l_p = 0.01\sigma$~\cite{lappala2013:macro, lappala2015:macro}, that is, almost zero bending stiffness. Once again, it is straightforward to extend the model to semiflexible chains in confinement, but we remain within the simplest possible polymer description.

 We adopted a simulation time-step of $\Delta t = 0.01\tau$. Considering that the  intrinsic energy of the system corresponds to $2.5$ kJ/mol in real units, the thermostat temperature will be about $300$~K. Taking the monomer size $\sigma$ to be around $0.3$ nm (typical size of an amino acid residue in proteins) and the average mass of an amino acid residue to be approximately $2\times 10^{-25}$ kg (with an average molecular weight of $\approx 110$), we can estimate that the LJ time unit becomes $\tau \approx 2$ ps. Consequently, by simulating $10^6 \Delta t$ time-steps in this coarse-grained approach, we can effectively trace the system dynamics over the $20$ ns period.

\subsection{The spring-wall model: chain between parallel walls}

\begin{figure}[tbp]
    \begin{center}
        \includegraphics[width=0.49\columnwidth]{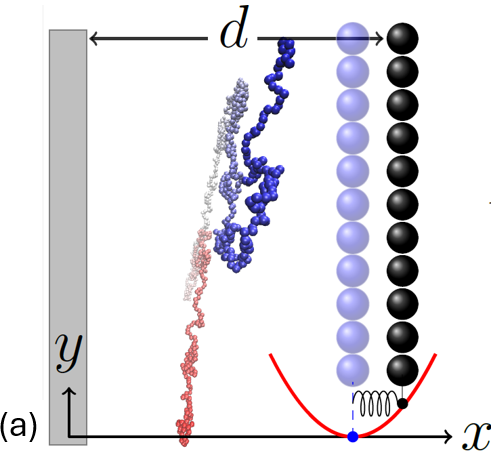}  
          \includegraphics[width=0.49\columnwidth]{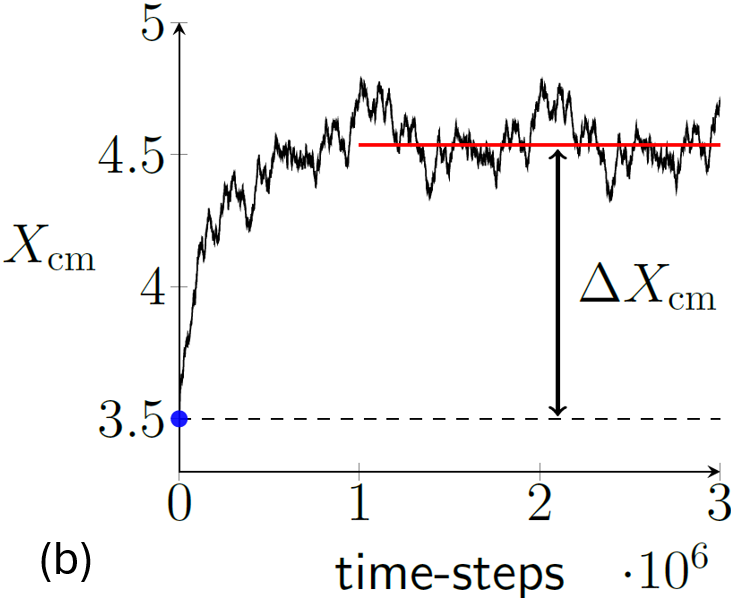}   
    \end{center}
    \caption{\label{fig:model}  (a) Two-dimensional schematic representation of a polymer chain confined between two walls: a reflective left wall and a spring-wall on the right. After equilibration, indicated by the black spring-wall, the equilibrium wall position, $\overline{X_{cm}}$, can be determined, allowing estimation of the average harmonic force and the equilibrium gap $d$. (b) Simulation result depicting the time evolution of the $x$-component of the center of mass of the wall, $X_{cm}$, for a self-avoiding chain with $N=1000$. The right wall is initially placed at $3.5\sigma$, corresponding to the minimum of the spring potential (blue bullet point). For visual reference, the snapshot of the initial configuration and a video of this simulation setup are provided in the SI.}
\end{figure}

The central point, and the purpose of this work, is to introduce a new method of `measuring' the entropic force exerted by the confined  chain. In  order to obtain the free energy of a  confined self-avoiding chain, we construct our confinement such as to empirically find the average force required to keep the a chain between parallel walls, one of which is movable -- itself constrained within a controlled harmonic potential, as depicted in Figure~\ref{fig:model}.
In this sketch, the polymer chain is confined along the $x$ direction by two parallel walls. The left wall is a traditional reflective wall, where the reflection is interpreted as the reversal of the perpendicular velocity component of a particle that moves towards the wall~\cite{lammps01,lammps02}. The right wall is made by a rigid plane of the same LJ particles, which is allowed to move along $x$ in a separate spring  potential. When the confined chain exerts an increasing force (pressure) on this wall -- it will move up the spring potential, and by measuring its average position we will directly measure the force. The plot in Figure~\ref{fig:model} illustrates this wall position stabilizing at a certain average value. 
Reflective walls are also employed in the $y$ and $z$ directions, but these boundaries are placed sufficiently wide to prevent the chain from coming into their proximity. Snapshots and a video of a simulation setup are provided in the Supporting Information.

The spring-wall model was constructed using the simulation resources provided within the LAMMPS framework~\cite{lammps01,lammps02}. The wall consists of $500^2$ LJ particles arranged in a  square lattice $(y,z)$ plane, with a lattice parameter of $a = 1.12\sigma$ (the minimum of the LJ potential).  Interactions among the wall particles are governed by the LJ potential, as given by Eq.~(\ref{eq:lj}), truncated at $r_{\mathrm{cut}}=3\sigma_{\mathrm{w}}$, where $\sigma_{\mathrm{w}} = 1.0$ and $\epsilon_{\mathrm{w}} = 5.0$.
Additionally, the interaction between the wall and the polymer chain is modeled using a repulsive LJ potential with $\sigma_{\mathrm{wch}} = 1.0$ and $\epsilon_{\mathrm{wch}} = 1.0$.

In order to attach the wall to a spring, we employ the LAMMPS routine \textsc{fix spring tether}. This command essentially applies a one-dimensional spring potential to the center of mass of a group of particles (the wall particles in our case)~\cite{lammps01,lammps02}. Further information about LAMMPS implementation can be found in the  Supporting Information.


The wall particles are initially placed at  the minimum of the spring potential, $X_0$, as illustrated in Figure~\ref{fig:model}. The stiffness of the spring is determined by a constant $k$, which gives the spring force acting on the center of mass of the wall~\cite{lammps01,lammps02}. For our purposes of capturing the equilibrium wall position and extracting the force exerted by the chain on the wall, we select a value of $k = 50$ in LJ potential units ($\epsilon/\sigma^2$).
However, it is important to appreciate that the entropic force exerted on the wall by the confined chain must be independent of a particular value of the spring-wall constant $k$, as long as it is sufficiently high to accurately capture the average harmonic force required to confine the polymer chain. This is verified in the  Supporting Information.

On allowing the polymer chain confined within our spring-wall model to equilibrate, we observe that the moveable wall is pushed outwards in the $x$-direction due to the force exerted by the chain. To measure this force, we output the time evolution of the $x$-component of the center of mass of this moving wall, denoted as $X_{cm}$. After reaching equilibrium, when the force generated by the chain is equal to the spring restoring  force, we are able to  estimate the average force exerted on the wall, as illustrated in Figure~\ref{fig:model}.

In this manner, we measure the equilibrium (average) wall position, $\overline{X}_{cm}$, represented by the red horizontal line in Figure~\ref{fig:model} (right), and thus compute the average harmonic force:
\begin{equation} \label{eq:force1}
 f  = k\Delta X_{cm}         = k(\overline{X}_{cm} - X_0),
\end{equation}
which is equal to the force exerted by the chain on the wall. However, a small caveat of this method is that we do not have a prescribed confinement gap $d$. Instead, we measure the  equilibrium gap $d$ between the walls in the same simulation. Since the fixed wall on the left is placed at $x=0.5\sigma$, the equilibrium gap is defined as $d = \overline{X}_{cm} - 0.5\sigma$. By obtaining the measured chain force $f$ as a function of the measured confinement length $d$, for the given chain length $N$ and the normalized temperature $k_B T/\epsilon$, we are able to analytically reconstruct the confinement free energy expression for a self-avoiding chain, as will be shown in the following section.

\section{Results and Discussion}

In order determine the confinement free energy of a self-avoiding chain, we first estimate the average harmonic force required to confine the polymer chain, as described by Eq.~(\ref{eq:force1}).
Initially, we simulated a polymer chain with $N$ particles in a good solvent, confined between two reflective walls at a distance of $2\sigma$, for a total of $10^7$ time-steps for a chain with $N=1000$ beads, and $2\times10^8$ time-steps for longer chains with $N=2000$ and $3000$ beads. The final configuration from this simulation is then used as the initial configuration for our well-equilibrated polymer chain in the subsequent simulations in our spring-wall model.

We conducted different simulations of a chain composed of $N$ beads confined within our spring-wall model. In each simulation, we varied the initial gap between the walls by adjusting the starting position of the spring wall, $X_0$. Through these simulations, we calculated the average harmonic force (normalized force), $f/k$, and the equilibrium gap, $d$, as presented for $N=1000$ in Table~\ref{tab-sarw}. Note that a very wide initial gap, $X_0 = 20.5\sigma$, did not yield any confinement force.

 \begin{table}[t]
 \caption{\label{tab-sarw} Results for a self-avoiding chain consisting of $N=1000$ beads.  All length parameters are given in units of $\sigma$.  Note that the error in the measurement of $\overline{X}_{cm}$ is approximately $\pm 0.10$.}
 \begin{tabular}{c c c c}
 \hline
 \hline
 $X_0$ & $\overline{X}_{cm}$ & $\Delta X_{cm} = \overline{X}_{cm}-X_0 $ & $d$\\
 \hline
 $3.50$	& $4.58\pm0.09$  & $1.08\pm 0.09$ & $4.08\pm 0.09$ \\
 $4.50$	& $5.31\pm0.09$  & $0.81\pm 0.09$ & $4.81\pm 0.09$ \\
 $6.00$	& $6.53\pm0.10$  & $0.53\pm 0.10$ & $6.03\pm 0.10$ \\
 $7.50$	& $7.84\pm0.10$  & $0.34\pm 0.10$ & $7.34\pm 0.10$ \\
 $9.00$	& $9.23\pm0.10$  & $0.23\pm 0.10$ & $8.73\pm 0.10$ \\
 $10.50$& $10.67\pm0.10$ & $0.17\pm 0.10$ & $10.17\pm 0.10$ \\
 $20.50$& $20.49\pm 0.10$ & $-0.01\pm 0.10$ & $19.99\pm 0.10$ \\
  \hline
  \hline
 \end{tabular} 
 \end{table}

 \begin{figure}[htbp]
    \begin{center}
        \includegraphics[width=0.9\columnwidth]{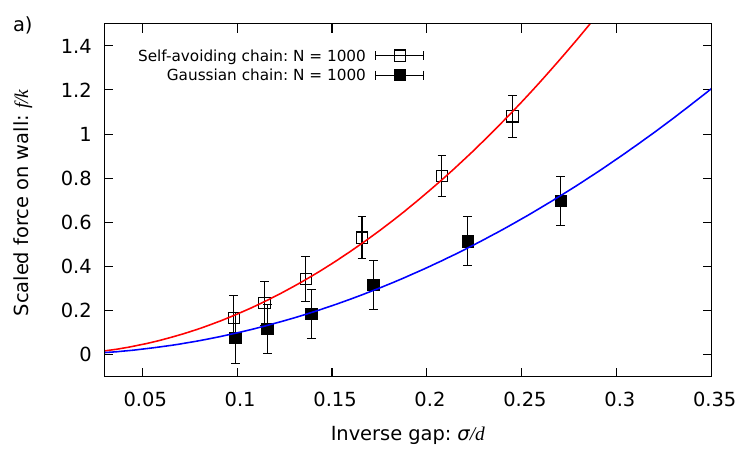}  
          \includegraphics[width=0.9\columnwidth]{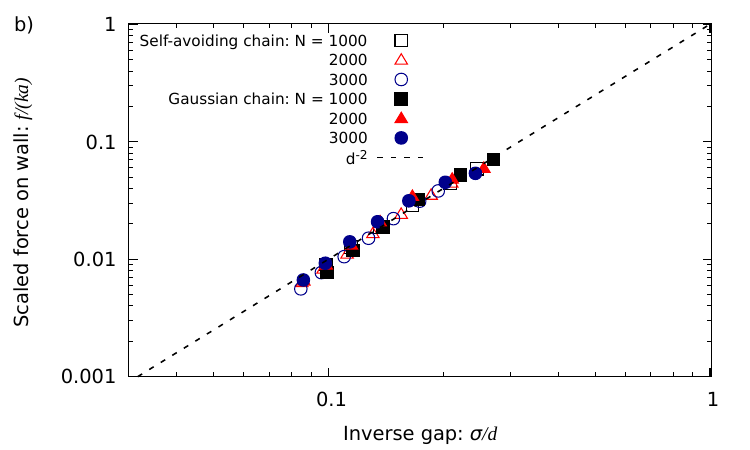}   
    \end{center}
      \caption{\label{fig:dgap} a) Normalized force, $f/k$, as a function of the inverse normalized gap, $\sigma/d$, for both Gaussian and self-avoiding chains with $N=1000$ beads (points). The y-error bars represent the standard deviations of average harmonic force. The solid lines represent the fitting curve for the quadratic  scaling law $f/k = a (\sigma/d)^2$ for both chains, where $a$ is a fitting parameter. In b), we present the force for both chains, normalized by the product $ka$. The data for the  self-avoiding chain are shown with empty points, and the Gaussian chain with filled points, for $N=1000$, $2000$ and $3000$ beads.}
\end{figure}

 Figure~\ref{fig:dgap} a) shows the normalized force, $f/k$, as a function of the inverse normalized gap, $\sigma/d$, obtained from our simulations for a self-avoiding chain with $N=1000$ beads (points). The solid line represents the fitted curve for the quadratic scaling law $f/k = a (\sigma/d)^2$, where $a$ is a fitting parameter. 
 As can be seen  the force exerted on the wall scales with the inverse  gap as $1/d^{2}$.
 
For comparison, we also present the data for a Gaussian (ideal) chain under similar conditions. Interestingly, here we find the same power-law dependence, indicating exactly the same repulsive force scaling with the confinement length. This implies that the equilibrium (entropic) free energy of the chain in this confinement scales as $F \propto 1/d$, a result not predicted or observed by any previous theories or simulations.

We also found this power-law force dependence for longer chains ($N=2000$ and $N=3000$), as shown in Figure~\ref{fig:dgap} b), where we present the force for both chains, normalized by the product $ka$. As can be seen,  all data are well described by a quadratic behavior ($1/d^{2}$). We also note small deviations for the first two points, where the gap is the widest and the chains are less confined, and the fluctuations are more pronounced. The values of $a$ and the individual force plots for the longer chains are provided in the Supporting Information.

 \begin{figure}[htbp]
    \begin{center}
        \includegraphics[width=0.9\columnwidth]{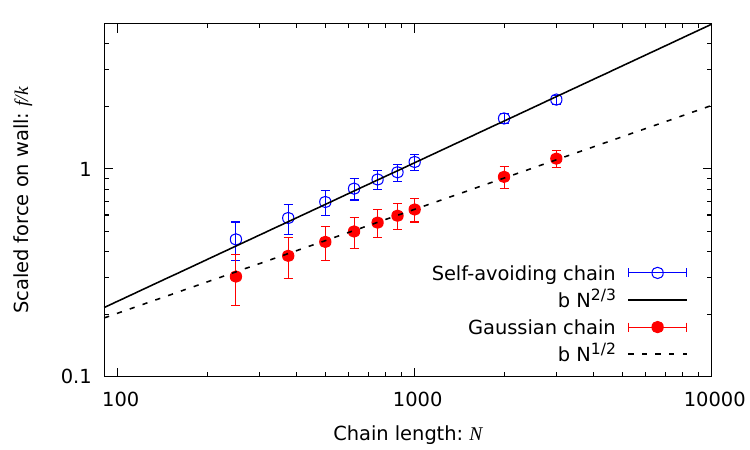}
    \end{center}
    \caption{\label{fig:Nplot} Log-log plot of the normalized force, $f/k$, as a function of chain length, $N$, for confined self-avoiding and Gaussian chains, with an initial gap between the walls equal to $3\sigma$. The y-error bars represent the standard deviations of the average harmonic force. The data for the self-avoiding chain, shown with empty points, were fitted to the function $f/k = bN^{2/3}$, represented by a solid line, while the data for the Gaussian chain, shown with filled points, were fitted to the function $f/k = bN^{1/2}$, represented by a dashed line.}
\end{figure}

\begin{figure}[htbp]
    \begin{center}
        \includegraphics[width=0.9\columnwidth]{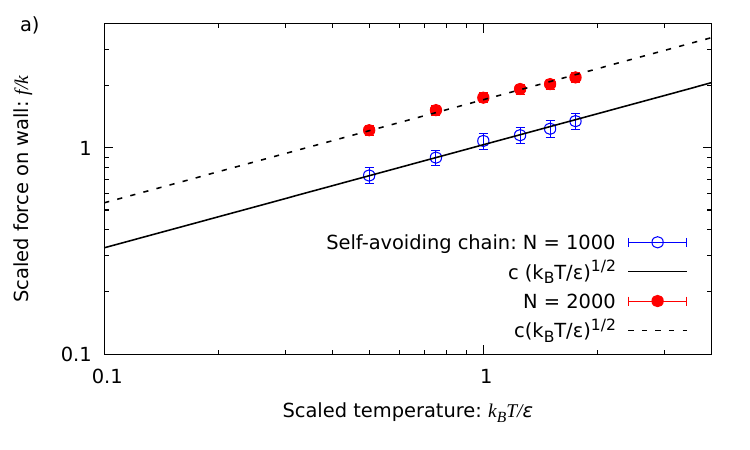}  \\
        \includegraphics[width=0.9\columnwidth]{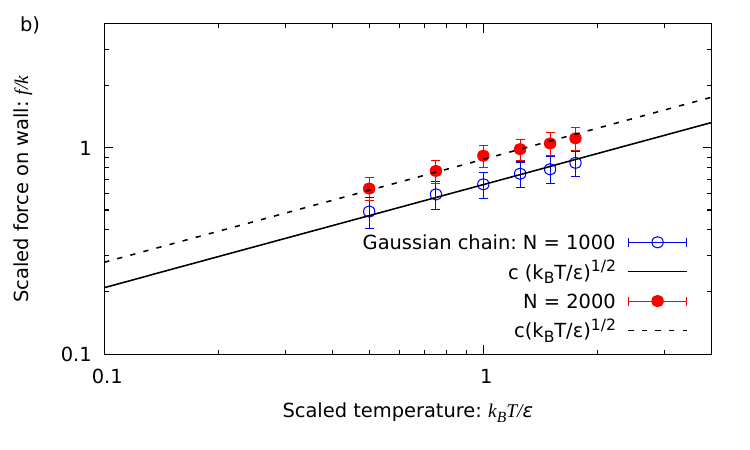}   
    \end{center}
      \caption{\label{fig:Tplot} Log-log plot of the normalized force, $f/k$, as a function of the scaled temperature, $k_BT/\epsilon$, for confined self-avoiding chains in a), and for Gaussian chain in b),  with the initial gap between the  walls equal to  $3\sigma$. The lines representing the fitting curves.}
\end{figure}

In order to derive the full expression for the force exerted by the self-avoiding chain on the wall, we  explore the power-law dependencies of the force with respect to both chain length, $N$, and the normalized temperature, $k_B T/\epsilon$. To achieve this, we conducted a series of simulations involving chains confined between two parallel walls, with the initial gap between the walls set to  $3\sigma$.

Remarkably, the result of this gives the normalized force $f/k$ with a clear scaling behavior with the chain length $N$. This scaling can be approximated by a power-law relationship, specifically $f \sim N^{2/3}$ for the self-avoiding chain (for the ideal Gaussian chain the similar scaling result was: $f \sim N^{1/2}$), as depicted in Figure~\ref{fig:Nplot}. Again, none of the previous theories or simulations have seen or suggested that. For instance, the minimization of the Flory model in Eq.\eqref{eq:flory} gives $F_\mathrm{eq} = 2k_BT \, N^{1/2} (\sigma/d)^{1/2}$, but this is for the self-avoiding chain. The classical Edwards and Freed model \cite{edwards1969} was for the ideal chain, and it gives $F_\mathrm{eq} = (\pi^2/6)k_BT \, N (\sigma/d)^{2}$.

Additionally, we also examine the effect of varying the normalized temperature, $k_B T/\epsilon$, and calculated the force exerted on the wall for both self-avoiding and Gaussian chains, as shown in Figure~\ref{fig:Tplot} (a) and (b), respectively, for a  fixed values of $N=1000$ and $2000$ monomers. In this case, the force is found to scale with the normalized temperature as $f/k = c(k_B T/\epsilon)^{1/2}$, with $c$ being an adjustable fitting constant.

In this manner, we are finding that the force exerted by the chain on wall is given by a factorized scaling expression: $ f  = f(E)f(N)f(d)$, where the separate contributions are due to the energy $E$, chain length $N$, and the confinement gap $d$, respectively. All were empirically determined as a power-law scaling relations from fittings in  Figs. \ref{fig:dgap}-\ref{fig:Tplot}. The full expression for the entropic force combines into:
\begin{equation} \label{eq:eqforce2}
 \begin{split}
    f & = k c   \left( \frac{k_B T}{\varepsilon}\right)^{1/2} \times k b N^{2/3} \times k a \frac{\sigma^2 }{d^2}       \\
        & = k^3 a b c \times  {\varepsilon}^{-1/2}({k_B T})^{1/2}N^{2/3} \frac{\sigma^2 }{d^2},
\end{split}
\end{equation}
where the constants $a$, $b$ and $c$  are the fitting parameters in Figs. \ref{fig:dgap}-\ref{fig:Tplot}, with their values  listed in Table~\ref{tab:abc} for both self-avoiding and Gaussian chain. The reader concerned that what we assert to be an entropic force appears not to be linearly proportional to $k_BT$ here should hold their doubt till the Eq.\eqref{eq:force} below, where we complete the dimensional analysis incorporating all the constants in their proper scaling forms. 

As previously mentioned, the force exerted by the chain on the wall is independent of the specific value of the spring constant $k$ (see SI for further details). However,  the product of the  $a$, $b$, and $c$ determines the magnitude of the force. For instance, from Table~\ref{tab:abc}, one can infer that the force increases with the number of monomers, $N$. Additionally, the force exerted by a self-avoiding is greater than that of a Gaussian chain,  as can also be obseved in Figs. \ref{fig:dgap}-\ref{fig:Tplot}. It is also important to note that the expression for the force a Gaussian chain is similar to the Eq.~(\ref{eq:eqforce2}), but differs only in the value of the numerical factor $abc$ and the power law contribution due to the chain length, which follows $N^{1/2}$.

\begin{table}[t]
 \caption{\label{tab:abc}  The values of the fitting parameters $a$, $b$ and $c$ obtained from Figs. \ref{fig:dgap}-\ref{fig:Tplot}  and  numerical factor $abc$ that appears in Eq.(\ref{eq:eqforce2}).}
\resizebox{\columnwidth}{!}{
    \begin{tabular}{cccccc}
        \hline
        \hline
        Chain  & N & a & b & c & $abc$   \\
            \hline
        \multirow{2}{*}{Self-avoiding}               & $1000$ &  $18.3\pm0.2$ & $0.0107\pm0.0001$ & $1.04\pm0.01$ &  $0.204\pm0.004$   \\
        & $2000$ &  $39.3\pm0.5$ & $0.0107\pm0.0001$ & $1.71\pm0.02$ &  $0.72\pm0.01$    \\
        \hline
    \multirow{2}{*}{Gaussian}              & $1000$ &  $9.5\pm0.3$  & $0.0203\pm0.0001$ & $0.66\pm0.01$ &  $0.13\pm 0.01$ \\ 
            & $2000$ &  $15.4\pm0.7$ & $0.0203\pm0.0001$ & $0.88\pm0.01$ &  $0.27\pm0.01$    \\
        \hline
        \hline
        \end{tabular}
    }
   \end{table}

 Importantly, in order to maintain dimensional consistency for the force, the product of the fitting parameters should have the dimensions of $[\sigma^5/\epsilon^2]$ in terms of LJ parameters. Furthermore, remembering that  $k = 50$ $\varepsilon/\sigma^2$ and  $\epsilon = k_{\mathrm{B}} T$ (see the Computational Details), we obtain that the force for the flexible self-avoiding chain confined between parallel walls  scales like:
\begin{equation} \label{eq:force}
 \begin{split}
    f & \sim \left (\frac{\varepsilon}{\sigma^2} \right)^3 \frac{\sigma^5}{\varepsilon^2}   {\varepsilon}^{-1/2}({k_B T})^{1/2}N^{2/3} \frac{\sigma^2 }{d^2} \\
     & \sim \frac{\varepsilon^{1/2}}{\sigma}({k_B T})^{1/2} N^{2/3} \frac{\sigma^2 }{d^2} \ = \  \mathcal{A}  \,  k_B T \, \frac{\sigma  N^{2/3}}{d^2}, \\
    \end{split}
\end{equation}
where the universal numerical constant was estimated from fitting the equation to computational data (see Supporting Information for detail): $\mathcal{A} =1/4$.
The corresponding free energy expression is derived by integrating the force in Eq. \eqref{eq:force} over the confinement length $d$. As a result, we obtain the confinement free energy analytically (which in effect means we have produce an  interpolation formula):
\begin{equation}\label{eq:free-energy}
\Delta F = \frac{1}{4} k_B T \frac{\sigma N^{2/3}}{d}.
\end{equation}

In the similar way, from the comparison results presented in Figs. \ref{fig:dgap}-\ref{fig:Tplot} for an ideal Gaussian chain, implementing the constraints on dimensionality, and estimating the remaining universal numerical constant from re-fitting the data (see Supporting Information), we conclude that the confinement free energy takes the form 
\begin{equation}\label{eq:free-energyG}
\Delta F = \frac{1}{3}  k_B T  \frac{\sigma N^{1/2}}{d}.
\end{equation}
The only fundamental difference between the two chain models is found in the power-law contribution due to the chain length $N$, which is mainly manifested across  the unconfined direction (parallel to the walls), representing chains with different equilibrium sizes.

To better understand the difference in flattened chain conformation in these two models, we computed the $x$-component of the radius of gyration, defined as the mean square $Rg_\mathrm{x}^2 = \frac{1}{N} \sum_{\mathrm{i}=1}^N (x_\mathrm{i} - x_{\mathrm{cm}})^2$, 
for chains with $N=1000$ monomers. Here, $x_i$ represents the x-coordinate of the position of the $i$th bead along the polymer chain, while $x_{\mathrm{cm}}$ denotes the x-coordinate of the center of mass of the polymer chains (along the confinement direction). 

\begin{figure}
    \begin{center}
       \includegraphics[width=0.9\columnwidth]{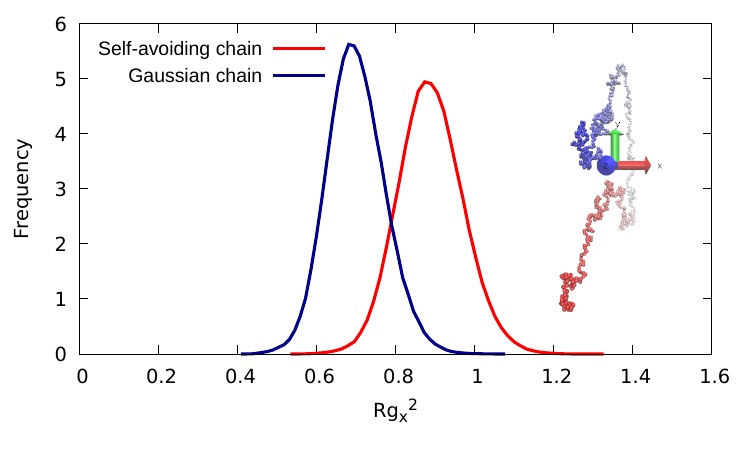}  
    \end{center}
       \caption{\label{fig:dist} Distribution of the $x$-component of the radius of gyration, $Rg_x^2$, for self-avoiding and ideal Gaussian chains, obtained by histogramming the values of $Rg_x^2$ (x-axis) from a simulation with  the initial wall gap of $3\sigma$. The inset represents a snapshot of an equilibrated self-avoiding chain confined along $x$. }
\end{figure}

In this manner, we show  the distribution of $Rg_\mathrm{x}^2$ in Figure~\ref{fig:dist}. Each distribution curve was obtained from a simulation  with a very narrow  initial wall gap of $3\sigma$.  It is important to remind the reader that the measured final equilibrium gap, $d$, is larger for the self-avoinding chain ($d=4.08\sigma$) compared to the ideal chain  ($d=3.70\sigma$), which explains why the average value $\langle Rg_x^2 \rangle$ is greater for the self-avoiding chain.
Additionally, we found that both chains are likely distributed as  Gaussian along the $x$-direction. This suggests that the self-avoiding and ideal chains are behaving similarly in the confinement direction, which means that their main differences are manifested across the unconfined region (parallel to the walls).

Therefore, the tighter parallel confinement enforces the ideal chain to be more stretched laterally, increasingly resembling an excluded-volume behavior in 2D. Since the majority of the chain conformations is explored in this lateral plane, we find this might explain why for both types of chains (ideal and self-avoiding), the equilibrium free energy exhibits the same scaling dependence on the confinement length, $1/d$.

\section{Conclusion}

In this study, we have introduced a new approach to empirically determine the confinement free energy  of polymer chains confined between parallel walls  through Brownian dynamics simulations. This approach is highly generic, and its concept can be used for many different types of chains, and types of confinement, as long as the moveable walls controlled by a spring force are properly constructed in the simulation. 

 {Notably, we found that both the ideal chain and the self-avoiding chain show their confinement free energy scaling with the gap as $1/d$, and accordingly the repuslive force (disjoining pressure) scaling as $1/d^2$. This is consistent with established scaling in the weak confinement regime \cite{gorbunov1995,Teraoka96} -- but we find the same scaling for the Gaussian ideal chains as well as for the self-avoiding chains -- and also well into the regime where one expects the stong confinement (i.e. $d < Rg$).  This does not correspond to any of the previously known theoretical models~\cite{casassa1967,edwards1969,milchev1998polymer,hassager2010}. Supporting Information gives more detail to this crossover issue. The answer has to be that we never reach the real strong confinement regime in our simulation, even though the gap was as low as $d=0.08 R_g$ (see Supporting Information for more detail). To us, it says that there are additional numerical factors in the crossover position that demand it to be so much smaller. }
 
{Earlier computational studies have have observed the crossover into the strong confinement regime. For instance, the method of `confinement analysis from bulk structures'  (CABS)~\cite{Wang2010,Muralidhar2014} has demonstrated a very clear crossover  near $d \approx R_g$ for different  chains types. The fundamental difference of our approach is that we do not use any theoretical input in measuring the confinement force, instead measuring it empirically, while the other methods -- from CABS to PMF-based methods -- do rely on heavy theoretical background: from calculating partition coefficients to various ways of statistical sampling.  We believe that future work with much longer chains is necessary to fully capture the crossover and the strong confinement scaling. It should be also mentioned that, in our approach, we encountered practical limitations that prevented us from reducing the gap further, as it automatically adjusts during the simulation.}
 We note that the parallel confinement enforces even the ideal the chain to stretch more in the lateral plane, resembling an excluded-volume behavior, which may explain the same $1/d$ dependence of the free energy for both models.

Similarly, the scaling with the chain length ($N^{1/2}$ for ideal chain and $N^{2/3}$ for self-avoiding chain) are unexpected and do not match well with many earlier studies. Nevertheless, we believe these results are `more correct' than others because we do not introduce any models or assumptions.  {Our method cannot claim to capture the full configurational entropy loss associated with confinement, but it does measure what the chain empirically exerts on the walls. In contrast, PMF-based methods (e.g. histogramming or umbrella sampling) reconstruct the free energy landscape and are more directly tied to the statistical mechanical definition of free energy.  It would be very interesting to compare the two methods for the same chain kind and conditions to better understand this conundrumof the disjoining pressure behaving is such a way.} 

 Using our method, we show how to construct the interpolated analytical expression for the free energy of a confined  chain to complement (or challenge) the classical theory of polymers in confined space. This simple concept could be beneficial also to model polymers under internal confinement as encountered in applications such as filled polymer or nanocomposites.

\section*{Supplementary Material}
The online Supporting Information document gives more details on technical aspects of this simulation. That includes the details of LAMMPS implementation for our spring-wall constraint, the evidence of how the disjoining pressure depends on the gap (and does not depend on the wall-spring constant $k$, and the explanation of our estimate of the numerical constant in the interpolated free energy.   \\

\section*{Acknowledgement}
This work was supported by the European Research Council H2020 AdG No: 786659, co-funding the visit of MSGF to Cambridge,  the Coordena\c c\~ao de Aperfei\c coamento de Pessoal de N\'ivel Superior, Brasil (CAPES), Fin. Code 001, and  São Paulo Research Foundation (FAPESP) grant number \#2023/03658-9. {The authors acknowledge the computational resources from the Darwin Supercomputer of the University of Cambridge High Performance Computing Service, and  the Santos Dumont Supercomputer (LNCC/MCTI, Brazil).} \\
\vspace{2 mm}
All data is available on request from the authors.

\section*{References}
%

\end{document}